\documentclass[letter,pdftex,twocolumn,epjc3]{svjour3}          

\RequirePackage[T1]{fontenc}
\usepackage[utf8]{inputenc}
\smartqed  

\RequirePackage{graphicx}
\RequirePackage{mathptmx}      
\RequirePackage{flushend}
\RequirePackage{amssymb}
\RequirePackage{amsmath}
\RequirePackage{amsfonts}
\RequirePackage{lineno}
\RequirePackage[numbers,sort&compress]{natbib}

\journalname{Eur. Phys. J. C}

\begin{document}

\title{Looking at the Axionic Dark Sector with ANITA}

\author{I. Esteban\thanksref{e1,addr1}
        \and
        J. Lopez-Pavon\thanksref{e2,addr2}
        \and
        I. Martinez-Soler\thanksref{e3,addr3,addr4,addr5}
        \and
        J. Salvado\thanksref{e4,addr1}
}

\thankstext{e1}{e-mail: ivan.esteban@fqa.ub.edu}
\thankstext{e2}{e-mail: jacobo.lopez@uv.es}
\thankstext{e3}{e-mail: ivan.martinezsoler@northwestern.edu}
\thankstext{e4}{e-mail: jsalvado@icc.ub.edu}

\institute{
  Departament de Fis\'ica Qu\`antica i Astrof\'isica and Institut de Ci\`encies del Cosmos, Universitat de Barcelona, Diagonal 647, E-08028 Barcelona, Spain\label{addr1}
  \and
  Instituto de F\'isica Corpuscular, Universidad de Valencia and CSIC, Edificio Institutos Investigaci\'on, Catedr\'atico Jos\'e Beltr\'an 2, 46980 Spain\label{addr2}
  \and
  Theoretical Physics Department, Fermi National Accelerator Laboratory, P.O. Box 500, Batavia IL 60510, USA\label{addr3}
  \and
  Department of Physics and Astronomy, Northwestern University, Evanston, IL 60208, USA\label{addr4}
  \and
  Colegio de F\'isica Fundamental e Interdisciplinaria de las Am\'ericas (COFI), 254 Norzagaray street, San Juan, Puerto Rico 00901\label{addr5}
}

\date{}

\maketitle

\begin{abstract}
  The ANITA experiment has recently observed two anomalous events
  emerging from well below the horizon. Even though they are consistent
  with tau cascades, a high-energy Standard Model or Beyond the Standard
  Model explanation is challenging and in tension with other
  experiments. We study under which conditions the reflection of generic
  radio pulses can reproduce these signals. Furthermore, we propose that
  these pulses can be resonantly produced in the ionosphere via
  axion-photon conversion. This naturally explains the direction and
  polarization of the events and avoids other experimental bounds.
\end{abstract}

\section{Introduction}

ANITA (ANtarctic Impulsive Transient Antenna) is a flying radio antenna
dedicated to measuring impulsive radio signals in the Antarctica~\cite{Gorham:2006fy, Gorham:2008dv, Hoover:2010qt}. In particular,
it can trigger pulses
originated by cosmic ray air showers~\cite{Hoover:2010qt}. 
ANITA has a very good angular resolution and is able to discern
whether the events are direct or reflected in the ice by measuring the
polarization and phase (so-called \emph{polarity} by the ANITA collaboration)
of the radio pulse. Two of the direct cosmic ray events observed in the first and third flights, which seem to be originated well below
the horizon ($27^\circ$ and $35^\circ$ respectively)~\cite{Gorham:2016zah,Gorham:2018ydl}, are particularly
intriguing and cannot in principle be interpreted as caused by high-energy cosmic rays. The only
standard model (SM) particle that could potentially traverse a large amount of
Earth matter (in this case around $6000$ and $7000$ km) and initiate a
particle cascade leading to these events is a very high-energy 
($\mathcal{O}({\rm EeV})$) neutrino. However, for these extremely high
energies, the neutrino-nucleon cross section leads to a very small
survival probability ($\lesssim 10^{-6}$) over the chord length of the
events, rendering such interpretation strongly
disfavored~\cite{Romero-Wolf:2018zxt,Fox:2018syq}. 

Recently, two potential SM explanations have been proposed in terms of
transition radiation~\cite{deVries:2019gzs} and reflection on
anomalous sub-surface structures~\cite{Shoemaker:2019xlt}. In both
cases, the origin of the anomalous events would be a reflected cosmic ray shower. Unless reflection occurs on a rather tilted surface, this
hypothesis is in principle in 2.5$\sigma$ tension with the observed
polarization angle of the first event~\cite{Gorham:2016zah}. Interestingly, both explanations predict
particular signatures. On the one hand, transition radiation predicts
that events with large elevations will be anomalous, in slight tension
with current data.  On the other hand, sub-surface structures predict
some amount of double events, which should also be generated by the calibration 
pulses emitted by the HiCal antenna~\cite{Gorham:2017xbo, Prohira:2019gos}. Therefore, dedicated searches and/or
more exposure are required to validate these possibilities.

Several Beyond the 
SM (BSM) scenarios have also been proposed to explain the origin
of these events in terms of high-energy
particles~\cite{Cherry:2018rxj,Anchordoqui:2018ucj,Huang:2018als,Yin:2018yjn,Fox:2018syq,Collins:2018jpg,Chauhan:2018lnq,Heurtier:2019git,Hooper:2019ytr,Cline:2019snp}. 
However, they are rather in tension with IceCube and Auger bounds~\cite{Cline:2019snp}.

In this Letter, we propose a novel origin for these intriguing
signals. We will show that reflected radio waves 
tend to present the 
properties of the mysterious ANITA events. 
Furthermore, we propose
that the radio signal is generated via the conversion of an axion-like
pulse. For the masses suggested by the data, this transformation happens
to be resonant in the Earth ionosphere. This process involves very
soft ($\mathcal{O}\left(\mu\mathrm{eV}\right)$) physics, invisible to
IceCube and Auger, that ANITA can potentially test with a dedicated
analysis.

\section{Anomalous ANITA events}

Atmospheric cosmic ray showers produce radio pulses with linear polarization perpendicular to the Earth magnetic field $\vec{B}^\oplus$, and a well-determined phase that flips at reflection. Since the magnetic field in the Antarctica is mostly vertical, ANITA searches for high-energy cosmic ray showers looking for horizontally polarized radio signals. In particular, the anomalous events are mostly horizontally polarized, and their phase led ANITA to interpret them as generated by up-going cosmic rays.

Any reflected electromagnetic wave, though, also tends to be horizontally polarized. In addition, if its origin is not a high-energy cosmic ray shower, its phase depends on the production mechanism and can thus match the one of the anomalous events.  
In this section, we will thus explore generic down-going radio waves reflected in the Antarctic ice as the origin for these events.

\begin{figure}[hbtp]
    \centering
    \includegraphics[width=0.48\textwidth]{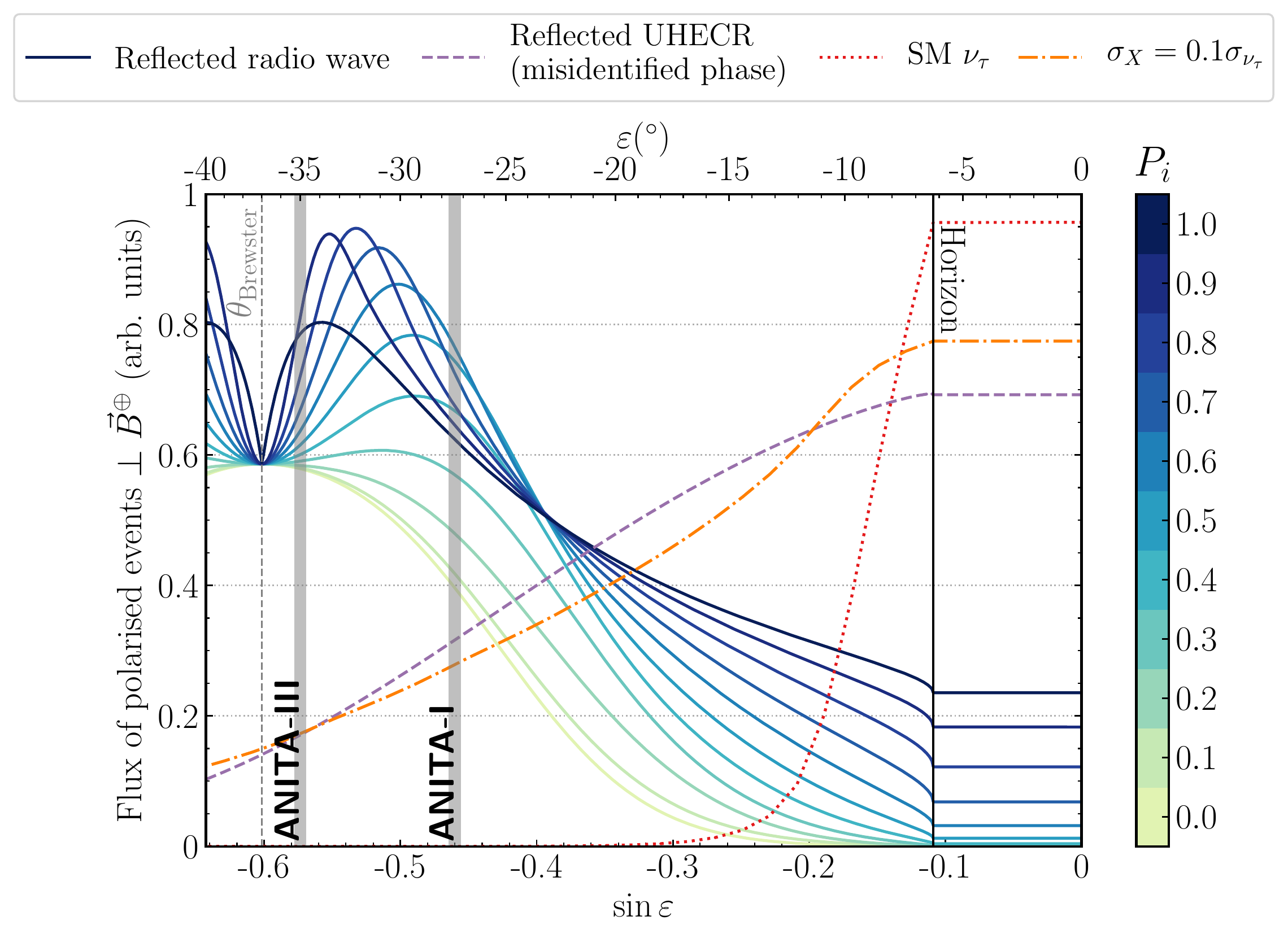}
    \caption{
    Expected angular distribution of linearly polarized events perpendicular to $\vec{B}^\oplus$ under different hypotheses as labelled in the legend (see the main text for further details). More information about the computation of the reflected radio wave expected distribution (solid lines) is given in the Supplemental Material. The two ANITA
      anomalous events are shown in gray. The black line marks the 
      elevation of the horizon, as seen from the ANITA balloon. The
      elevation range plotted corresponds to ANITA's angular acceptance~\cite{Gorham:2008dv}.
      }
    \label{fig:flux}
\end{figure}

This hypothesis is illustrated in Fig.~\ref{fig:flux}, where we show in solid the expected angular distribution of reflected events perpendicular to $\vec{B}^\oplus$ as a function of their elevation $\varepsilon$. We have assumed an incident isotropic flux, linearly
polarized with random polarization angles and a given degree of
polarization $P_i$. The reflected flux peaks at
elevations where light reflects close to the Brewster angle $\theta_B
\sim 53^\circ$ (corresponding to $\varepsilon \sim
-37^\circ$)\footnote{The maximum is not exactly at $\theta_B$ because
the Earth magnetic field has a small horizontal component, and
therefore ANITA searches for events that are slightly tilted with
respect to the horizontal.}, defined as the angle at which the reflected
signal is polarized exactly in the horizontal direction. The
elevations of the observed events, within $1 \sigma$, are shown in
gray; they are both close to the peak.

In Fig.~\ref{fig:flux} we also show the expected fluxes associated to
other relevant hypotheses for the origin of the anomalous events.  
First, a SM tau neutrino flux (dotted line) which strongly peaks at the horizon and is therefore highly
disfavored. Next,  
a generic BSM high-energy particle with a nucleon
interaction cross section 10 times weaker than that of the SM
neutrino (dot-dashed). 
The latter hypothesis partially alleviates the tension in
the angular distribution, but tension with IceCube and Auger data
remains~\cite{Cline:2019snp}. Another possibility is that ANITA
misidentified reflected events originated by ultra-high-energy cosmic ray (UHECR) air showers, classifying them instead as direct
events. This hypothesis, shown by the dashed line, is disfavored
since it would require phase misidentification by ANITA, which is
excluded at $\sim 4.5\sigma$~\cite{Rotter:2017}. However, alternative
scenarios have been recently proposed to support this
possibility~\cite{deVries:2019gzs,Shoemaker:2019xlt}.

According to Fig.~\ref{fig:flux}, our hypothesis of reflected radio
waves, linearly polarized in random initial directions, explains the
up-going direction of the two events. The other alternative scenarios
shown in the figure predict more events close to the horizon,
and so they could be discriminated from our proposal as ANITA
accumulates more exposure.

\begin{figure*}[hbtp]
    \centering
    \includegraphics[width=\textwidth]{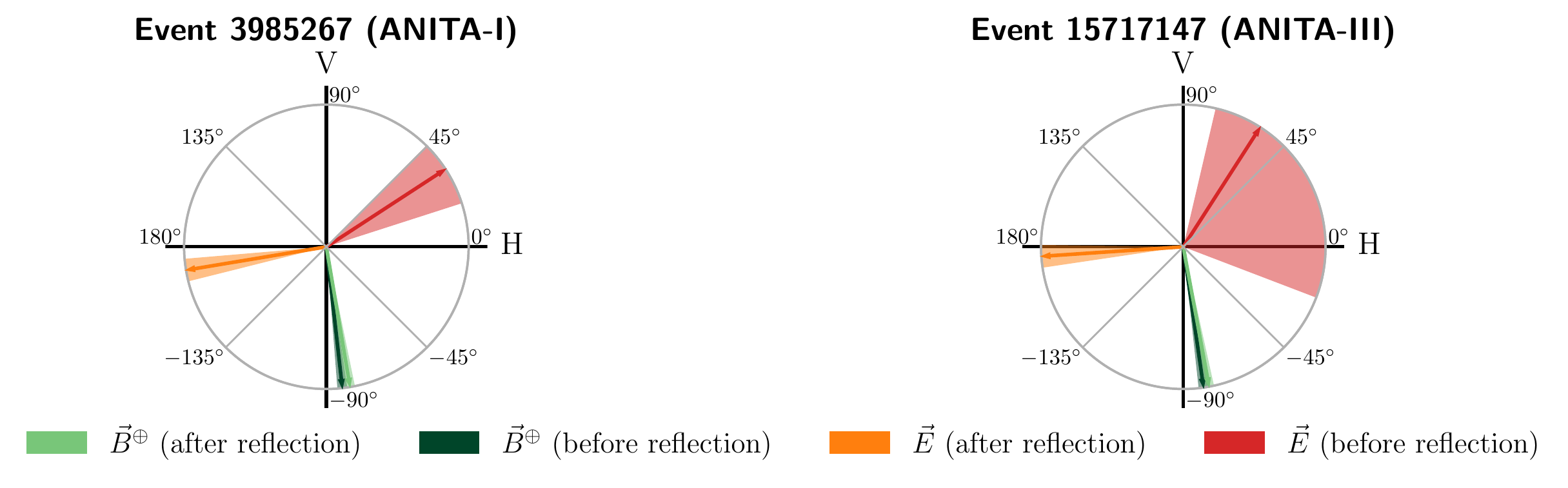}
    \caption{Observed (orange) and incident (red)
      polarization angles. We also show the projection of the Earth
      magnetic field~\cite{WMM2005,WMM2010}, in the reflected (light
      green) and incident (dark green) polarization planes. The shaded regions correspond to
$1\sigma$ uncertainties assuming a $4.6^\circ$ uncertainty in the
determination of the polarization direction~\cite{Gorham:2018ydl}, and a $2^\circ$
uncertainty in the orientation of the Earth magnetic
field~\cite{Rotter:2017}. }
    \label{fig:pol}
\end{figure*}

Another key observable in ANITA, not shown in Fig.~\ref{fig:flux}, is
the polarization angle.  In Fig.~\ref{fig:pol} we quantify under which
initial conditions our proposed hypothesis is able to reproduce not only the angular distribution, but also
the observed polarization angle of the anomalous events. 
We show the reflected
(orange) and incident (red) polarization angles for both
events assuming that the signal is fully polarized. For the signal to
be identified as an UHECR, the reflected (orange) electric field
should be orthogonal to the Earth magnetic field (green), i.e, it should 
essentially be along the horizontal direction (H) as it is shown in the figure.

Since both events emerge at elevations close to the Brewster angle $\theta_B$, the vertical (V) component
of the reflected electric field becomes quite suppressed. This is the
reason why the uncertainty on the incident polarization angle is
larger than the uncertainty on the reflected polarization angle\footnote{The relation between the incident and reflected polarization angles with respect to the horizontal, $\psi_i$ and $\psi_r$, is non-linear: $\psi_i = \arctan \left(\frac{r_H}{r_V} \tan \psi_r\right)$, where $r_H$ and $r_V$ are the horizontal and vertical Fresnel reflection coefficients~\cite{Goldstein:2003}. Thus, the uncertainty on $\psi_i$ is asymmetric. 
This asymmetry is particularly evident for $\psi_i \sim \pm 90^\circ$  and, thus, for ANITA-III since this event is closer to $\theta_B$ ($r_V \sim 0$).}. 
This
effect is particularly relevant for the second event, since
it is closer to $\theta_B$ and has a reflected vertical component
compatible with 0. The first event, on the other hand, has a non-zero
vertical component, which significantly tightens the range of allowed
incident polarization angles.

\section{Axion-like origin}

An incoming isotropic flux of linearly polarized radio waves with the
required initial conditions to reproduce the observed ANITA signals,
though, cannot in principle be explained in the SM. Below, we will construct a BSM explanation based on the
following requirements:
\begin{itemize}
    \item The source must generate a flux of impulsive
      radio signals, spatially isolated with a linear polarization and phase consistent
      with Fig.~\ref{fig:pol}.
    \item Due to the tension with IceCube and Auger data, the
      production process must not involve high-energy particle
      cascades.
\end{itemize}
The first requirement comes from the ANITA triggering system~\cite{Hoover:2017, Mottram:2017, Rotter:2017}, that requires a source of isolated impulsive signals. Notice that, even though it is not a necessary requirement, astrophysical sources are expected to be isotropically distributed, as assumed in Section 2.

The second requirement calls for a new physics mechanism able to
coherently generate electromagnetic waves. This phenomenon should
produce waves with frequencies $\sim \mathcal{O}(1 \,\mathrm{GHz})$,
i.e., it can be associated with very low energies $\sim
\mathcal{O}(10^{-7}\,\mathrm{eV})$. An archetypal BSM example are axion-like particles (ALPs), that convert into photons in the presence
of an external electromagnetic field.

The ALPs, first proposed to solve the Strong CP
problem~\cite{Peccei:1977hh,Weinberg:1977ma,Wilczek:1977pj}, arise in
many extensions of the SM and can 
constitute the dark matter in our
universe~\cite{Preskill:1982cy,Abbott:1982af,Dine:1982ah}.
Furthermore, they present self-interactions that produce a very rich
and complex phenomenology. In particular, different phenomena, like condensation
into a bosonic soliton or instabilities leading to scalar field
bursts, may produce impulsive, spatially localized configurations of
the scalar field~\cite{Ruffini:1969qy, Kolb:1993hw, Davidson:2013aba,
Chavanis:2011zi, Hertzberg:2016tal, Braaten:2015eeu, Eby:2015hyx,
Eby:2016cnq, Levkov:2016rkk, Chavanis:2017loo, Visinelli:2017ooc,
Chavanis:2018pkx, Braaten:2018nag, Amin:2019ums, Eby:2019ntd, Olle:2019kbo}. 
If an axion pulse, with a macroscopic occupation number, reaches us,
it can transform into the electromagnetic pulses observed by ANITA via the
interaction with the Earth magnetic field. 

For this phenomenon to explain the ANITA anomalous events, its rate should be $\sim \mathrm{month}^{-1}$. Any calculation of such rate is highly model dependent, as the phenomenology of non-linear ALP interactions is complex and still under study. Nevertheless, it is estimated that in a local neighbourhood $\sim 1 \, \mathrm{pc}^3$ there can be between $10^{10}$ and $10^4$ ALP overdensities~\cite{Fairbairn:2017sil}. Each overdensity would develop unstable bosonic solitons within time scales $\sim 10^{-2}$--$	10^7$ years~\cite{Levkov:2018kau, Levkov:2016rkk}. Thus, the required rate to explain the anomalous events could plausibly be attained.

An ALP is a pseudo-scalar field, $a$, that interacts with photons via a Lagrangian density $\frac{1}{4} g_{a \gamma \gamma} a F_{\mu \nu} \tilde{F}^{\mu \nu}$, where $F_{\mu \nu}$ is the electromagnetic field tensor and $\tilde{F}_{\mu \nu}$ its dual.
In presence of an external magnetic field $\vec{B}^\oplus$, the classical equations of motion for the axion $a$ and
electric field $\vec{E}$ in a plasma with free electron density $n_e$ are
given by~\cite{Raffelt:1987im, Sikivie:1983ip}
\begin{equation}
    \left[\partial_z^2\! +\! \omega^2\! +\! 
    \begingroup 
    \setlength\arraycolsep{1pt}
    \begin{pmatrix}
    -\omega_p^2 & -i \omega_p^2 \frac{\Omega_z}{\omega} & 0 \\
    i \omega_p^2 \frac{\Omega_z}{\omega} & - \omega_p^2 & 
    - g_{a \gamma \gamma} B^\oplus_y \omega^2 \\
    0 & -g_{a \gamma \gamma} B^\oplus_y  & -m_a^2
    \end{pmatrix}\endgroup \right]\!
    \begin{pmatrix}
    E_x \\ E_y \\ a
    \end{pmatrix}\!\! =\! 0 \, ,
    \label{eq:masterEq}
\end{equation}
where we have assumed waves propagating in the $Z$ direction, a
magnetic field in the $YZ$ plane, and we have taken the Fourier
transform in time. Here $\omega_p = \sqrt{\frac{e^2}{m_e} n_e}$ is
the plasma frequency, $\Omega_z = \frac{e B^\oplus_z}{m_e}$ is the
cyclotron frequency of the plasma, $m_e$ is the electron mass, and $e$ is the 
electron charge in natural units. 
Notice that the axion mass $m_a$ in the above equations can include extra contributions depending on the particular scalar self interactions considered. 

When the Faraday rotation effects (non-diagonal terms proportional
to $\Omega_z$) are switched off, the solution of these equations gives an electromagnetic wave linearly polarized
parallel to the external magnetic field. The Faraday rotation 
then leads to the rotation of this polarization in the $XY$ plane.

\section{Resonance in the ionosphere}

Eq.~\eqref{eq:masterEq} is a system of coupled wave equations with two
characteristic frequencies, $\omega^2 - \omega_p^2$ and $\omega^2 -
m_a^2$. When both frequencies are equal, axions convert resonantly
into photons~\cite{Raffelt:1987im,Yoshimura:1987ma,Sikivie:1983ip}.  
The smallest observed frequency of the anomalous events, $\omega_\mathrm{min} \sim 0.25 \, \mathrm{GHz}$, requires $m_a \lesssim \omega_\mathrm{min} \lesssim 10^{-7} \, \mathrm{eV}$. These masses happen to be in the range of typical
$\omega_p$ values of the Earth ionosphere~\cite{Kelley:2009}, and thus
an incoming axion pulse with $m_a \lesssim 10^{-7} \,\mathrm{eV}$
would cross a region where $m_a \simeq \omega_p$, resonantly
transforming into electromagnetic waves.
This resonant conversion takes place in a relatively narrow region
$\mathcal{O}(10\, \mathrm{km})$. The produced radio pulse will initially
be polarized parallel to the Earth magnetic field, with a phase that
depends on the axion wave phase and the sign of the axion-photon
coupling $g_{a\gamma\gamma}$.  This signal will later traverse the
rest of the ionosphere, rotating its polarization vector due
to the Faraday effect.

Generically, the axion burst will cross the resonant region twice, producing two radio pulses.
Since the propagation through the ionosphere
partially unpolarizes and decoheres the radio pulse generated during
the resonance, there is a trade-off between having enough ionosphere
to generate the Faraday rotation required to match the signals (see
Fig.~\ref{fig:pol}) and keeping the pulse coherent\footnote{We have checked, for different
ionospheric configurations, that the resonantly generated pulses can remain
coherent enough while undergoing sufficient Faraday rotation.}. These effects are
mainly controlled by the density of free electrons $n_e$ in the
ionosphere, which fluctuates in time in more than one order of
magnitude~\cite{Kelley:2009}. Therefore, we can distinguish different
phenomenological scenarios depending on the values of $n_e$. 
For small values, 
the pulse produced in the second resonant region will not undergo enough Faraday
rotation, producing a mostly vertically polarized (parallel to
$\vec{B}^\oplus$) signal. This pulse would be triggered out by the ANITA
analysis or strongly suppressed due to the reflection close to
$\theta_B$. However, the pulse generated during the first resonance
would potentially be detectable.
For high values of $n_e$,  
the
first resonantly produced pulse would become incoherent, but the one
generated in the second resonance would still be coherent and also
experience enough Faraday rotation to pass ANITA's triggers. 

\begin{figure}[hbtp]
    \centering
    \includegraphics[width=0.48\textwidth]{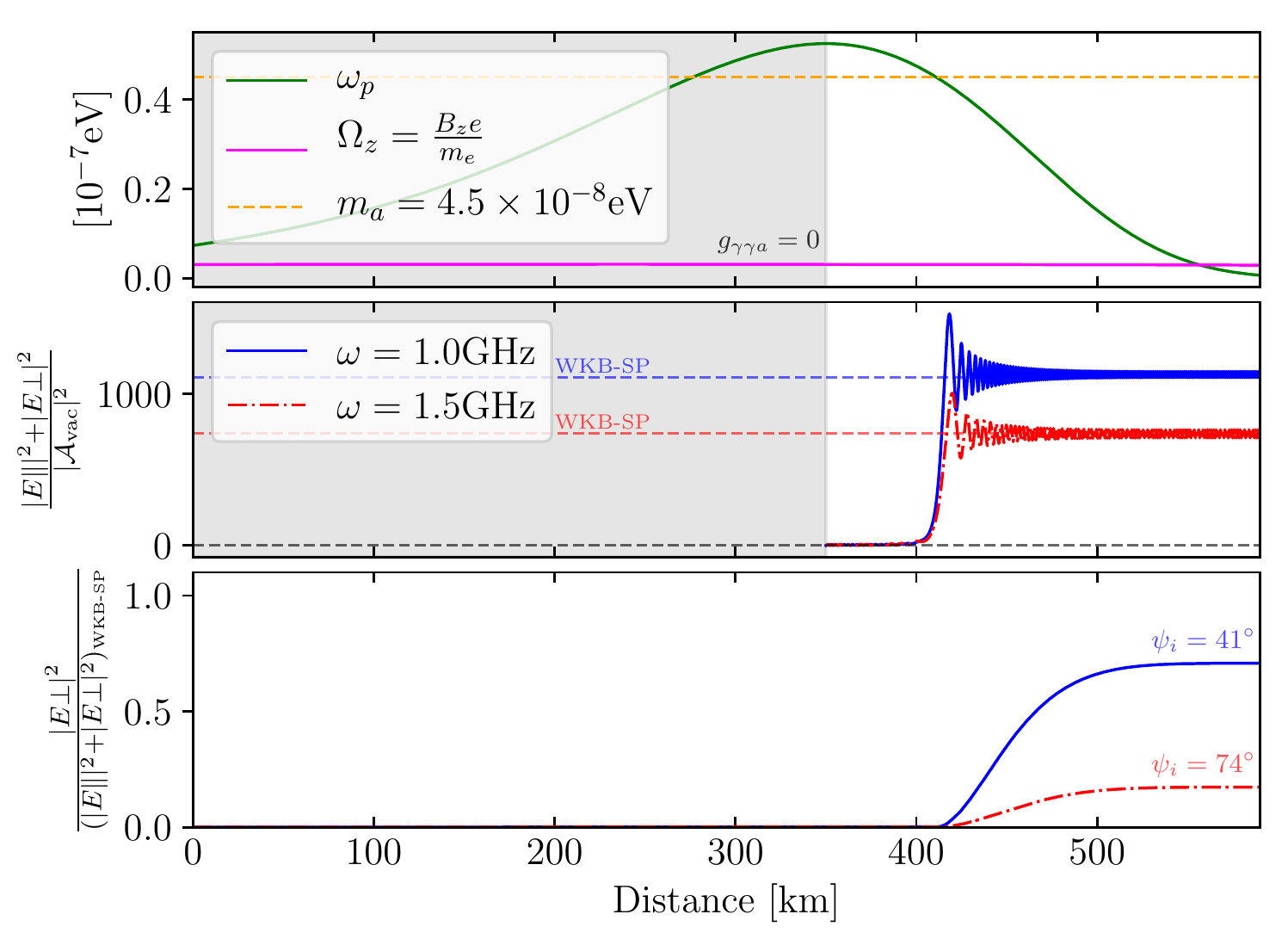}
    \caption{\textit{Top panel:} $\omega_p$ and $\Omega_z$ profile in the ionosphere, assuming
      a Chapman layer profile~\cite{Kelley:2009, Chapman_1931} with a plausible maximum free electron density $n_e^\mathrm{max} = 2 \times 10^6 \, \mathrm{cm}^{-3}$, along with the axion mass chosen in the
      simulation. \textit{Central panel:} electric field squared amplitude,
      normalized to the vacuum axion-photon conversion squared
      amplitude. In dashed, we show the result using the analytical
      approximation given by Eq.\eqref{eq:WKB}. The dashed gray line corresponds to the vacuum conversion squared amplitude.
      In the shaded gray region, we have switched off the axion-photon
      coupling in order to show only the propagation of the second
      resonant burst. \textit{Bottom panel:} squared component of the electric field perpendicular to the Earth magnetic field, generated via Faraday rotation. We have normalized it to the total
      squared amplitude as given by Eq.\eqref{eq:WKB}.}
    \label{fig:secondbang}
\end{figure}

To illustrate these effects, in Fig.~\ref{fig:secondbang} we show the numerical solution of Eq.~\eqref{eq:masterEq} for the
second resonant conversion, which takes place at the lowest
altitudes (details on how Eq.~\ref{eq:masterEq} is numerically solved can be found in Appendix C).

The top panel shows the plasma
frequency $\omega_p$ and the cyclotron frequency $\Omega_z$\footnote{For the considered Earth magnetic field model~\cite{WMM2005,WMM2010}, $B^\oplus_y/B^\oplus_z\approx 2 $.} as a function of the propagated distance in the
ionosphere, together with the axion mass $m_a$ value considered; the resonance occurs when both lines coincide. The
central panel shows the total squared amplitude of the electric field due to
resonant axion-photon conversion for different frequencies, normalized
to the vacuum squared amplitude $|\mathcal{A}_\mathrm{vac}|^2=\left(\frac{2 a_0 g_{a \gamma \gamma} B^\oplus_y \omega^2}
    {m_a^2}\right)^2$, where $a_0$ is the amplitude of the incoming axion field. Finally, the bottom panel shows
the squared projection of the electric field in the direction perpendicular to the Earth magnetic field, generated due to Faraday rotation.

The enhancement factor with respect to the vacuum axion-photon transition observed in Fig.~\ref{fig:secondbang} can be qualitatively understood using the WKB and stationary phase
approximations~\cite{Hook:2018iia} to solve 
eq.~\eqref{eq:masterEq}

\begin{equation}
    |E|^2 = |a_0|^2\left(\frac{2 g_{a \gamma \gamma} B^\oplus_y \omega^2}
    {m_a^2}\right)^2 \left[ \frac{\pi}{4} \frac{m_a^2/k^2}
    {\left.\frac{\mathrm{d} \omega_p^2/m_a^2}{k \, \mathrm{d} z}
    \right|_\mathrm{res}} \right] \, ,
    \label{eq:WKB}
\end{equation}
where $E$ is the electric field after the resonance, 
$k=\sqrt{\omega^2-m_a^2}$ is the wave number,   
and the derivative is evaluated at the resonance. The
term  
inside the square brackets gives the enhancement due to the resonance in the plasma. This approximate solution is shown by the dashed lines in Fig.~\ref{fig:secondbang}. In the ionosphere, $\omega_p$ varies over
distances $\mathcal{O}(10\, \mathrm{km})$, much longer than the
wavelengths that can be observed by ANITA,
$\mathcal{O}(\mathrm{m})$. Therefore, the denominator in Eq.~\eqref{eq:WKB} is rather small, leading to the $\mathcal{O}(10^2-10^3)$ global enhancement observed in Fig.~\ref{fig:secondbang}. As can be observed in the central panel of Fig.~\ref{fig:secondbang}, before reaching the resonance the squared amplitude of the electric field is basically equal to its value when the axion-photon conversion takes place in vacuum. That is, outside the resonant region, the axion-photon conversion rate is essentially the one in vacuum, different from zero but negligible compared with the resonant value.

The projection of the electric field shown in the bottom panels of
Fig.~\ref{fig:secondbang} is directly
related to the polarization angle of the radio pulse. 
When the pulse leaves the ionosphere, this angle $\psi_i$ must be consistent with the red regions in Fig~\ref{fig:pol}
for the ANITA anomalous events to be reproduced.
In particular, for $\omega=1 \,\mathrm{GHz}$  
we obtain a 
polarization angle of $\psi_i \approx 41^\circ$, in agreement with Fig.~\ref{fig:pol} and thus consistent 
with ANITA-I and ANITA-III.
For $\omega=1.5\, \mathrm{GHz}$  
we
have $\psi_i \approx 74^\circ$, consistent 
with ANITA-III but in disagreement with ANITA-I. This is
because in this case the Faraday rotation effect does
not generate enough horizontal (essentially orthogonal to
$\vec{B}^\oplus$) component of the electric field to reproduce the
first event. A different ionospheric profile and/or $m_a$, though, would in
general give a different result. Thus, pulses experiencing different levels of Faraday rotation, with diverse polarization directions after the ionosphere, can be generated.
 
\begin{figure}[hbtp]
    \centering
    \includegraphics[width=0.48\textwidth]{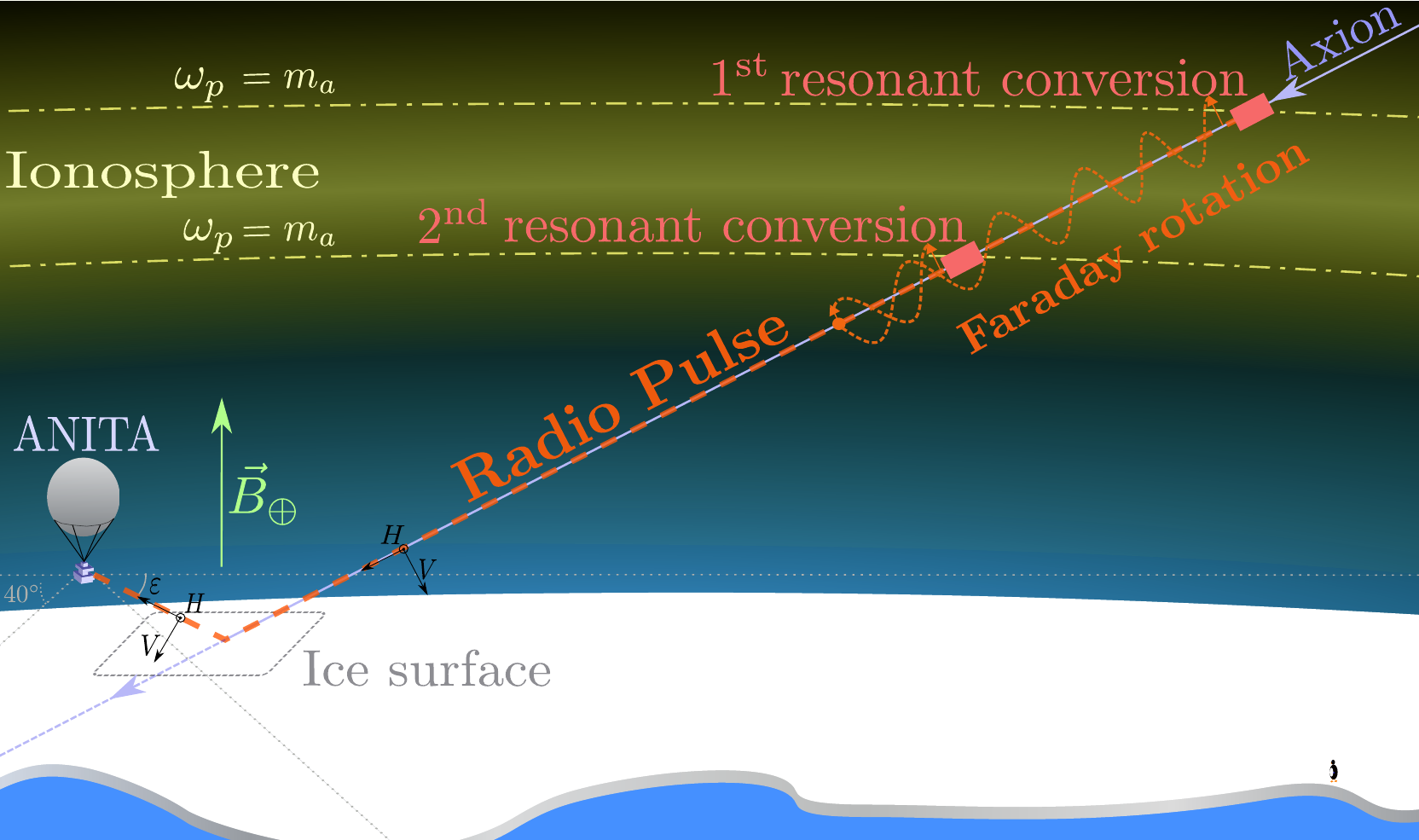}
    \caption{Sketch of an axion burst arriving to the ionosphere,
      undergoing resonant conversion, 
      Faraday rotation, and reflecting on the ice surface
      before reaching ANITA.}
    \label{fig:cartoon}
\end{figure}

In summary, the ANITA anomalous events could be due to an axion
burst that resonantly converts into photons in the Earth ionosphere 
matching the required conditions of the isotropic, linearly polarized flux previously discussed above and shown in Figs.~\ref{fig:flux} and \ref{fig:pol}. Our proposal is schematically summarized in
Fig.~\ref{fig:cartoon}.

\section{Spectral properties and ALP scenario}

The characteristics of the observed ANITA events can also be used to
extract properties of the ALP burst.

On the one hand, we can infer information about the frequency of the
burst. Both anomalous events show a large correlation with a
cosmic-ray template, and \cite{Gorham:2018ydl} shows the Amplitude
Spectral Density of the ANITA-III anomalous event.  We have checked that both
spectral requirements can be satisfied within experimental
uncertainties with a Gaussian pulse with central frequency $\omega
\lesssim 2.5\, \mathrm{GHz}$ and width $\sigma_\omega = 1.5-4\,
\mathrm{GHz}$.

\begin{figure}[hbtp]
    \centering
    \includegraphics[width=0.48\textwidth]{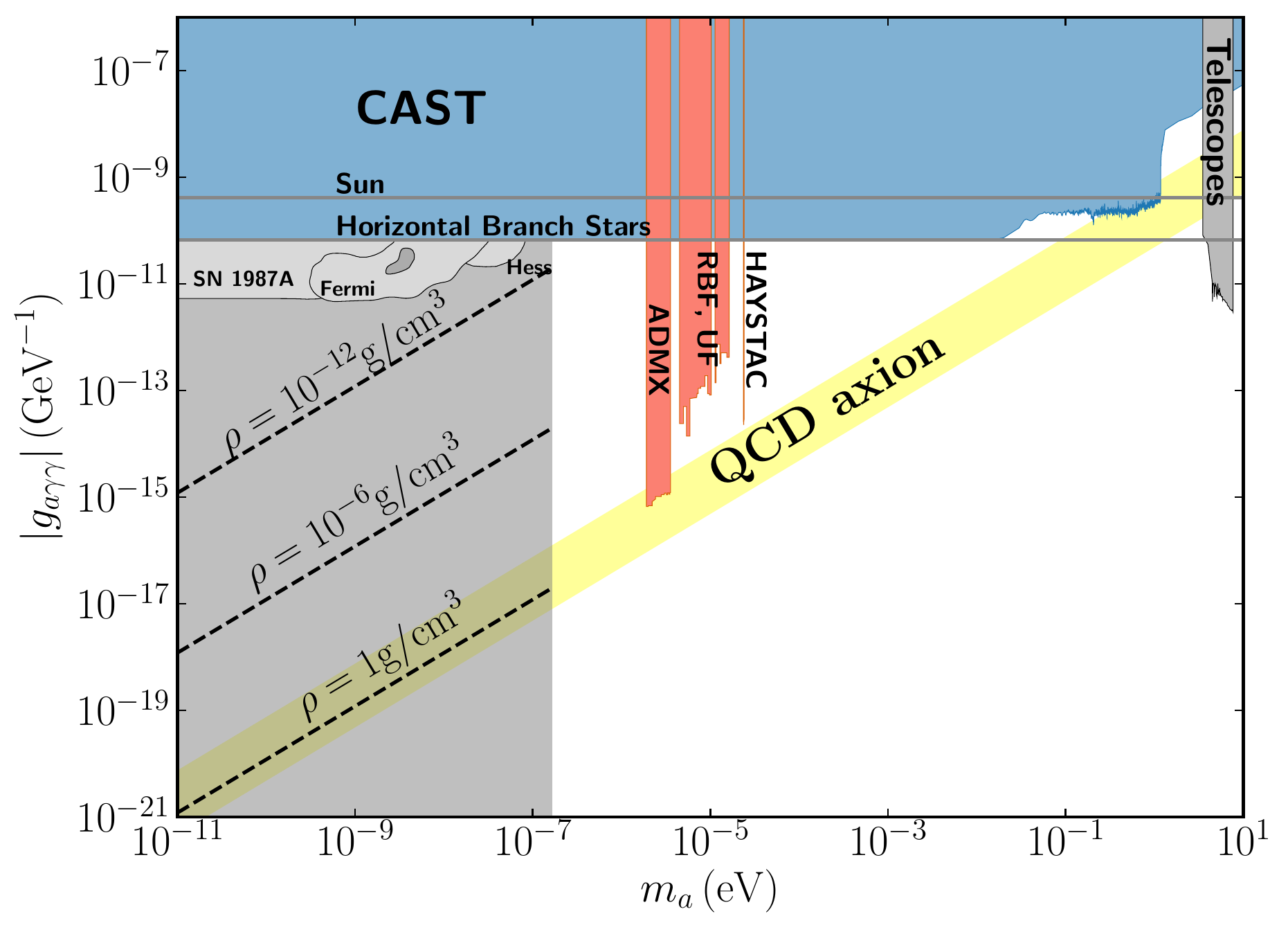}
    \caption{In gray, estimated axion mass $m_a$ and axion-photon coupling $g_{a \gamma \gamma}$ consistent with the ANITA events for different densities $\rho$ of the incoming axion burst. The minimum observed frequency $\omega \gtrsim 0.25 \, \mathrm{GHz}$ forces $m_a \lesssim 1.6 \cdot 10^{-7} \, \mathrm{eV}$. We also show current experimental constraints~\cite{Bahre:2013ywa,Armengaud:2019uso,Ayala:2014pea,Payez:2014xsa,Anastassopoulos:2017ftl,Tanabashi:2018oca}. The yellow region is compatible with the QCD axion models~\cite{DiLuzio:2016sbl}.}
    \label{fig:gvsma}
\end{figure}

On the other hand, we can also extract information on the mass $m_a$ and
coupling $g_{a \gamma \gamma}$ of the ALP. Using Eq.~\eqref{eq:WKB},
we can estimate the relation between the observed electric field
amplitude $\sim 1 \,\mathrm{mV/m}$, $m_a$,
$g_{a \gamma \gamma}$, and the amplitude of the axion field burst $a_0$.  The latter can, in turn, be determined by the energy
density of the burst $\rho \sim |a_0|^2 \omega^2$. Considering  
$B^\oplus \sim 0.45 \, \mathrm{G}$~\cite{WMM2005,WMM2010}, a wave frequency $\omega \sim 1.5\,\mathrm{GHz}$, attenuation due to reflection on the Antarctic ice, and typical resonance-enhancement factors
$\frac{\mathrm{d}\omega_p^2/m_a^2}{\mathrm{d}z} \sim 10^{-2} \, \mathrm{km}^{-1}$, we show in Fig.~\ref{fig:gvsma} an estimation for the values of $g_{a \gamma \gamma}$ and $m_a$ consistent with ANITA for different energy densities $\rho$ of the incoming axion burst. The gray region shows the mass-coupling range compatible with the minimum observed frequency $\omega \sim 0.25 \, \mathrm{GHz}$\, (corresponding to $m_a \lesssim 10^{-7} \, \mathrm{eV}$), and the dashed lines label different densities of the incoming axionic burst. For densities $\gtrsim 10^{-12} \, \mathrm{g}/\mathrm{cm}^2$ all present experimental bounds are evaded~\cite{Bahre:2013ywa,Armengaud:2019uso,Ayala:2014pea,Payez:2014xsa,Anastassopoulos:2017ftl,Tanabashi:2018oca}.

\section{Conclusions}

In this work, we have explored the directional and polarization
properties of the anomalous ANITA events 3985267 (ANITA-I) and
15717147 (ANITA-III). We have found that the reflection of an
isotropic flux, linearly polarized in arbitrary directions, can
naturally accommodate both observables. This is mostly due to the
triggering of ANITA, that favors horizontally polarized events,
together with reflection close to the Brewster angle.

Requiring a polarized flux not produced via high-energy cascades in order to avoid the IceCube and Auger bounds, we have proposed a generation mechanism based on
the axion-photon conversion in the Earth magnetic field of a
classical, high occupation number, axion burst.

Interestingly, we have also found that this conversion is dominated by
a resonance naturally occurring in the ionosphere for the radio
frequencies observed by ANITA.
After traversing the remaining part of the ionosphere, the signal will
undergo Faraday rotation, providing pulses polarized in different
directions that can explain the mysterious events.

Our proposal can already be tested reanalyzing the data collected
by ANITA, including the fourth flight data currently under
analysis. If the hypothesis presented in the second section is
correct, relaxing the triggering that requires geomagnetically
correlated events should reveal events with a non-suppressed vertical component emerging from angles different
from $\theta_B$. On the other hand, an axion-like origin generically
predicts two consecutive events with different polarizations and/or coherence,
since the axion burst experiences two resonant transitions into
photons along its propagation through the ionosphere. Extra signals
could thus be observed by searching for doubled events which may require
decreasing the coherence threshold. 

\begin{acknowledgements}
  
We would like to thank I. Est\'evez and N.P. Plaza for useful discussions
about polarimetry and geoscience. We also thank A. Caputo, P. Coloma, L. 
Molina Bueno and S. Witte for discussions and careful reading of the
manuscript. This work is supported by EU Networks\,\, FP10ITN\, ELUSIVES\,
(H2020-MSCA-ITN-2015-674896)\\
and INVISIBLES-PLUS
(H2020-MSCA-RISE-2015-690575), by the\\
MINECO grant
FPA2016-76005-C2-1-P and by the Maria de Maeztu grant MDM-2014-0367 of
ICCUB. JLP acknowledges support by the ``Generalitat Valenciana"
(Spain) through the ``plan GenT" program (CIDEGENT/2018/019). Fermilab
is operated by the Fermi Research Alliance,\,\, LLC\,\, under\, contract\,
No.\, DE-AC02-07CH11359\, with\, the\\
United States Department of
Energy. I.M.S. acknowledge travel support from the Colegio de Fisica
Fundamental e Interdisciplinaria de las Americas
(COFI). I.E. acknowledges support from the FPU program fellowship
FPU15/03697.
\end{acknowledgements}

\onecolumn
\appendix

\section{Expected elevation distribution of reflected events}

The ANITA anomalous events are compatible with being 100\% linearly
polarized perpendicular to the Earth magnetic
field~\cite{Gorham:2016zah, Gorham:2018ydl}. Therefore, in
Fig.~\ref{fig:flux} we have computed the expected distribution of
events that satisfy these properties within experimental
uncertainties.

To do so, we have considered a distribution of incoming radio pulses,
characterized by their incident angle $\theta_i$, polarization angle
$\psi_i$ and degree of polarization $P_i$. The pulses are assumed to
be uniformly distributed in $\theta_i$ and $\psi_i$, with constant
$P_i$. The Stokes formalism~\cite{Goldstein:2003} relates the
reflected and incident polarization states, and so it allows to obtain
the expected distribution of events as a function of the corresponding
reflected polarization angle, reflected degree of polarization, and
angle of reflection.

The reflected radio pulses should match the polarization properties
observed by the ANITA
collaboration~\cite{Gorham:2016zah,Gorham:2018ydl}. They have to be,
within experimental uncertainties, compatible with being observed as
100\% linearly polarized perpendicular to the Earth magnetic
field. Imposing this requirement, we obtain the expected distribution
of events as a function of the reflected angle (which is directly
related to the elevation $\varepsilon$). The resulting distribution is
what we show in solid lines in Fig.~\ref{fig:flux}.

In the computation, we have assumed an index of refraction for the
Antarctic surface of $n=1.35$~\cite{Gorham:2008dv,Gorham:2017xbo}.  In
addition, in order to impose the expected signal to be compatible with
being observed as 100\% linearly polarized perpendicular to the Earth
magnetic field, information about the experimental uncertainties on
the polarization angle $\psi$ and degree of polarization $P$ is
required. The former is reported by the ANITA collaboration to be
4.6$^\circ$~\cite{Gorham:2018ydl}, whereas to our knowledge the latter
is not available. In order to estimate this uncertainty, we have
considered that the degree of polarization is given by
      
\begin{equation}
  P = \frac{\sqrt{Q^2 + U^2 + V^2}}{I} \, ,
\end{equation}
which is determined by the Stokes parameters of the incoming radio pulse,
      
\begin{align}
  I & = \frac{1}{N} \sum_{i=1}^N |\varepsilon_H^i|^2 + |\varepsilon_V^i|^2 \, , \\
  Q & = \frac{1}{N} \sum_{i=1}^N |\varepsilon_H^i|^2 - |\varepsilon_V^i|^2 \, , \\
  U & = \frac{1}{N} 2 \mathrm{Re} \left[\sum_{i=1}^N \varepsilon_H^i \left(\varepsilon_V^i\right)^* \right] \, , \\
  V & = \frac{1}{N} 2 \mathrm{Im} \left[\sum_{i=1}^N \varepsilon_H^i \left(\varepsilon_V^i\right)^* \right] \, ,
\end{align}
where $\varepsilon \equiv E + i \hat{E}$, $\hat{E}$ is the Hilbert
transform of the electric field $E$ and the subindices $H$ and $V$
denote the horizontal and vertical components of the electric field at
$N = 500$ discrete time instants~\cite{Rotter:2017}.

According to Ref.~\cite{Rotter:2017}, the uncertainty on the Stokes
parameters (and so the uncertainty on $P$) is dominated by random
Gaussian white noise. We have thus assumed $\varepsilon^i$ to be
uncorrelated Gaussian random variables with a mean value given by the
true electric field and a standard deviation $\delta$. The
experimental uncertainties on the Stokes parameters (and therefore on
$P$) have been estimated by randomly generating $\varepsilon^i$. The
only free parameter, $\delta$, has been adjusted to reproduce ANITA's
uncertainty on the polarization angle reported
in~\cite{Gorham:2018ydl}.

Our estimation leads to an uncertainty on the degree of polarization
of $\sim 0.13$. In any case, we have checked that considering a
different input for this uncertainty does not change significantly our
conclusions.

Finally, for the SM $\nu_\tau$ hypothesis and the flux coming from a
generic particle with an interaction cross section 10 times smaller
than the SM $\nu_\tau$, we have simulated propagation through the
Earth with the numerical library
$\nu$-SQuIDS~\cite{Delgado:2014kpa,squids, nusquids}.

\section{Initial conditions for a partially polarized pulse}

\begin{figure*}[hbtp]
    \centering
    \includegraphics[width=\textwidth]{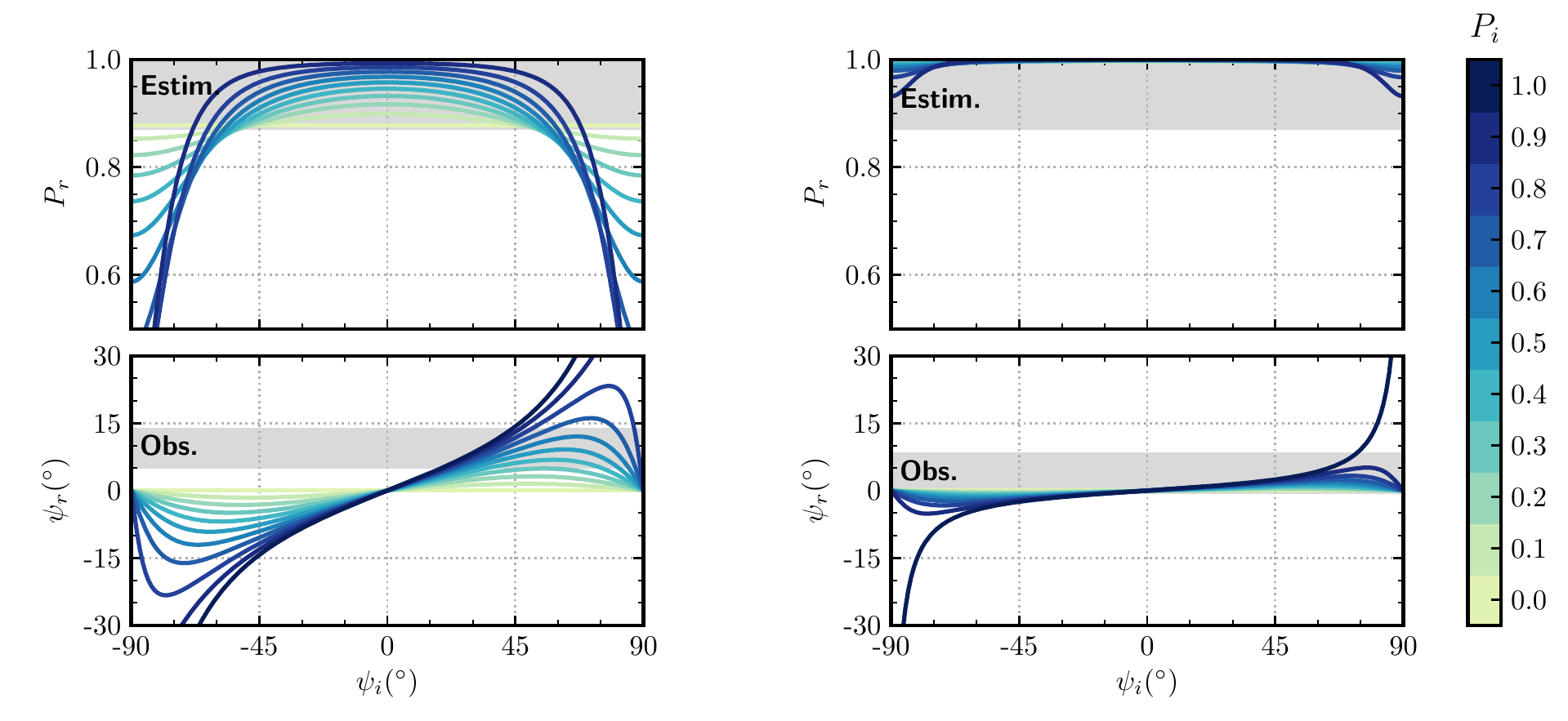}
    \caption{Reflected polarization angle $\psi_r$ and degree of
      polarization $P_r$ as a function of the incident polarization
      angle $\psi_i$ and degree of polarization $P_i$ (labeled by the
      color). The $1\sigma$ allowed region for the polarization angle~\cite{Gorham:2018ydl} (bottom panels) and our corresponding estimation for the degree of polarization (top panels) are shown in gray. All angles are measured with respect to the horizontal.}
    \label{fig:pol2}
\end{figure*}

The incident pulse does not need to be fully polarized. In the following, we generalize
Fig.~\ref{fig:pol} by relaxing this hypothesis. To
visualize its effect, in the top and
bottom panels of Fig.~\ref{fig:pol2} we 
show the relation among the reflected degree of polarization 
and polarization angle, $P_r$ and $\psi_r$; and the incident
degree of polarization and polarization angle, $P_i$ and $\psi_i$. The corresponding $1\sigma$ allowed region for $\psi_r$ extracted from~\cite{Gorham:2018ydl} is shown in gray in the bottom panels. The gray region in the top panels corresponds to our estimation for the degree of polarization's allowed region at $1\sigma$.

Comparing Fig.~\ref{fig:pol} to the bottom panels in Fig.~\ref{fig:pol2}, we conclude that the range of
allowed incident polarization angles increases once we slightly relax
the assumption of a fully polarized signal. This is because the
unpolarized part of the signal leads to a horizontally polarized
component after reflection, which tends to tilt the polarization angle
closer to the horizontal, i.e., closer to being perpendicular to $\vec{B}^\oplus$.
Comparing the ANITA-I and ANITA-III panels, we notice that the level
of initial polarization and range of values of the incident
polarization angle is less stringent for ANITA-III, as expected since
its incident angle is closer to $\theta_B$.

\section{Numerical Integration}

Eq.~\eqref{eq:masterEq} has fast oscillating solutions with very different time-scales:
the frequency of the waves, $\omega$, is much larger than all the other scales in the
system such as the plasma frequency, $\omega_p$; the axion mass, $m_a$; and the axion-photon coupling term, $g_{a \gamma \gamma} B^\oplus_y
\omega^2$. For this reason, the numerical integration is not straightforward: we have first written all fields $C(z)$ as $C(z) = \tilde{C}(z) e^{i k z}$, with $k = \sqrt{\omega^2 - m_a^2}$. In this way, the fast oscillations driven by $\omega$ are effectively separated from the axion-photon conversion and Faraday rotation effects, which take place with much longer characteristic times.

In our computation, the initial condition for Eq.~(\ref{eq:masterEq}) is a
burst of axion field and no electromagnetic field. Given the smallness of $g_{a\gamma\gamma}$, the feedback of the generated electromagnetic wave on the axion field can be neglected to a good approximation. Thus, the axion field amplitude can be considered constant. 
This allows us to simplify the problem and solve Eq.~(\ref{eq:masterEq}) for $E_i$ ($i=\{x,y\}$) considering the axion field as a constant source. 

Furthermore, any set of coupled second order differential equations can be written as a set of first order differential equations (at the cost of doubling the number of equations) and thus, separating the real and imaginary parts of the electric field $\tilde E_i=\Re (\tilde E_i) + i\Im (\tilde E_i)$ and denoting $\tilde D_i = \partial_z \tilde E_i$, we can rewrite Eq.~\eqref{eq:masterEq} as

\begin{equation}
  \partial_z
  \begin{pmatrix}
    \Re (\tilde{D}_x) \\ \Im (\tilde{D}_x)\\ \Re (\tilde{D}_y) \\ \Im (\tilde{D}_y) \\ \Re (\tilde{E}_x)\\ \Im (\tilde{E}_x)
    \\ \Re (\tilde{E}_y)\\ \Im (\tilde{E}_y)
  \end{pmatrix} =
  \begin{pmatrix}
    0&2k&0&0& \omega_p^2 - m_a^2 & 0 & 0 & -\omega_p^2 \frac{\Omega_z}{\omega} \\
    -2k&0&0&0&0& \omega_p^2 - m_a^2 & \omega_p^2 \frac{\Omega_z}{\omega} & 0 \\
    0&0&0&2k& 0 & \omega_p^2\frac{\Omega_z}{\omega} & \omega_p^2 - m_a^2 & 0 \\
    0&0&-2k&0& - \omega_p^2\frac{\Omega_z}{\omega}& 0 & 0 & \omega_p^2 - m_a^2 \\
    1&0&0&0&0&0&0&0\\
    0&1&0&0&0&0&0&0\\
    0&0&1&0&0&0&0&0\\
    0&0&0&1&0&0&0&0
  \end{pmatrix}
  \begin{pmatrix}
    \Re (\tilde{D}_x) \\ \Im (\tilde{D}_x)\\ \Re (\tilde{D}_y) \\ \Im (\tilde{D}_y) \\ \Re (\tilde{E}_x)\\ \Im (\tilde{E}_x)
    \\ \Re (\tilde{E}_y)\\ \Im (\tilde{E}_y)
  \end{pmatrix} + 
  \begin{pmatrix}
    0 \\ 0\\ g_{a \gamma \gamma} B^\oplus_y \omega^2 \tilde{a} \\ 0 \\ 0\\ 0 \\ 0\\ 0
  \end{pmatrix}.
  \label{eq:masterEqNum}
\end{equation}

Where $a$ is the amplitude of the axion pulse arriving to the Earth. This differential equation can be easily solved with a standard numerical integrator. In particular, we have used the one from the \texttt{scipy} python package, which is essentially a wrapper of the \texttt{FORTRAN} library \texttt{odepack}\cite{odepack}. We used the default integrator: an explicit Runge-Kutta method of order 5(4) featuring an adaptive step size, setting the precision goal to be unnoticeable in Fig.~\ref{fig:secondbang}.

\bibliographystyle{spphys}
\bibliography{biblio}

\end{document}